\documentclass[12pt]{article}
\newcommand\apjl{{ApJ}}%
\newcommand\apjs{{ApJS}}%
\newcommand\aap{{A\&A}}%
\usepackage{graphicx}
\usepackage{times}
\usepackage{geometry}
\usepackage[utf8]{inputenc}
\usepackage{enumitem,amssymb}
\usepackage{natbib}
\newlist{thematic}{itemize}{8}
\usepackage{pifont}

\begin{document}
\huge
\centerline{Astro2020 Science White Paper:}
\centerline{Low Mass Stars as Tracers of Star Formation }
\centerline{ in Diverse Environments}
\normalsize

\vskip 0.25 in

\noindent \textbf{Thematic Areas:} \hspace*{60pt} $\square$ Planetary Systems \hspace*{10pt} $\boxtimes$ Star and Planet Formation \hspace*{20pt}\linebreak
$\square$ Formation and Evolution of Compact Objects \hspace*{31pt} $\square$ Cosmology and Fundamental Physics \linebreak
  $\square$  Stars and Stellar Evolution \hspace*{1pt} $\square$ Resolved Stellar Populations and their Environments \hspace*{40pt} \linebreak
  $\square$    Galaxy Evolution   \hspace*{45pt} $\square$             Multi-Messenger Astronomy and Astrophysics \hspace*{65pt} \linebreak
  
\noindent
\textbf{Principal Author:}\\
\noindent
Name: S. Thomas Megeath \\
 \noindent
Institution: Dept. of Physics and Astronomy, University of Toledo \\
 \noindent
Email: tommegeath@gmail.com\\

\noindent 
\textbf{Co-authors:} Marina Kounkel (WWU), Stella Offner (U. of Texas), Rob Gutermuth (U. of Massachusetts), Hector Arce (Yale), Will Fischer (STScI), Zhi-Yun Li (U. of Virginia), Sarah Sadavoy (CfA), Ian Stephens (CfA), John Tobin (NRAO), Elaine Winston (CfA)\\


\noindent
\textbf{Abstract - Background:} low-mass stars are the dominant product of the star formation process, and they trace star formation over the full range of environments, from isolated globules to clusters in the central molecular zone.  In the past two decades, our understanding of the spatial distribution and properties of young low-mass stars and protostars has been revolutionized by sensitive space-based observations at X-ray and IR wavelengths. By surveying spatial scales from clusters to molecular clouds, these data provide robust measurements of key star formation properties.   \\

\noindent
{\bf Goal:} with their large numbers and their presence in diverse environments, censuses of low mass stars and protostars can be used to measure the dependence of  star formation on   environmental properties, such as the density and temperature of the natal gas, strengths of the magnetic and radiation fields, and the density of stars. Here we summarize how such censuses can answer three basic questions: i.) how is the star formation rate influenced by environment, ii.) does the IMF vary with environment, and iii.) how does the environment shape the formation of bound clusters?  Answering these questions is an important step toward understanding star and cluster formation across the extreme range of environments found in the Universe.  \\

\noindent
{\bf Requirements:} sensitivity and angular resolution improvements will allow us to study the full range of environments found in the Milky Way. High spatial dynamic range ($< 1''$ to $> 1^{\circ}$ scales) imaging with space-based telescopes at X-ray, mid-IR, and far-IR  and ground-based facilities at near-IR and sub-mm wavelengths are needed to identify and characterize young stars. Wide field proper motion studies at 2~$\mu$m can be used to identify members and study kinematics. Multi-object and IFU near-IR spectroscopy are needed to measure the ages and masses of young stars and measure radial velocities. Finally, simulations are essential to investigate the physical mechanisms present in different environments and how those mechanisms will affect observed correlations.


\newpage

Wide field surveys for low-mass pre-ms stars and protostars can now study star formation across entire molecular clouds and star formation complexes, i.e. 100~pc scales, encompassing the diverse environments found within each cloud \citep[e.g.,][]{2009ApJS..181..321E,2012AJ....144..192M,2015ApJ...802...60K}. These surveys can detect the full range of young ($< 5$~Myr) stellar objects in nearby ($\le 500$~pc) clouds, from protostars, to pre-ms  stars with disks, to young stars without optically thick disks  \citep[e.g.,][]{2007prpl.conf..361A,2014prpl.conf..195D}. In the next decade, we will extend these studies to more distant, diverse environments, using the following techniques to survey the full range of star formation found in the Milky Way (see Fig.~1): 

\begin{figure}[b!]
\begin{minipage}[c]{0.5\textwidth}\center{\includegraphics[width = 3.6 in, trim =  0 0 0 20]{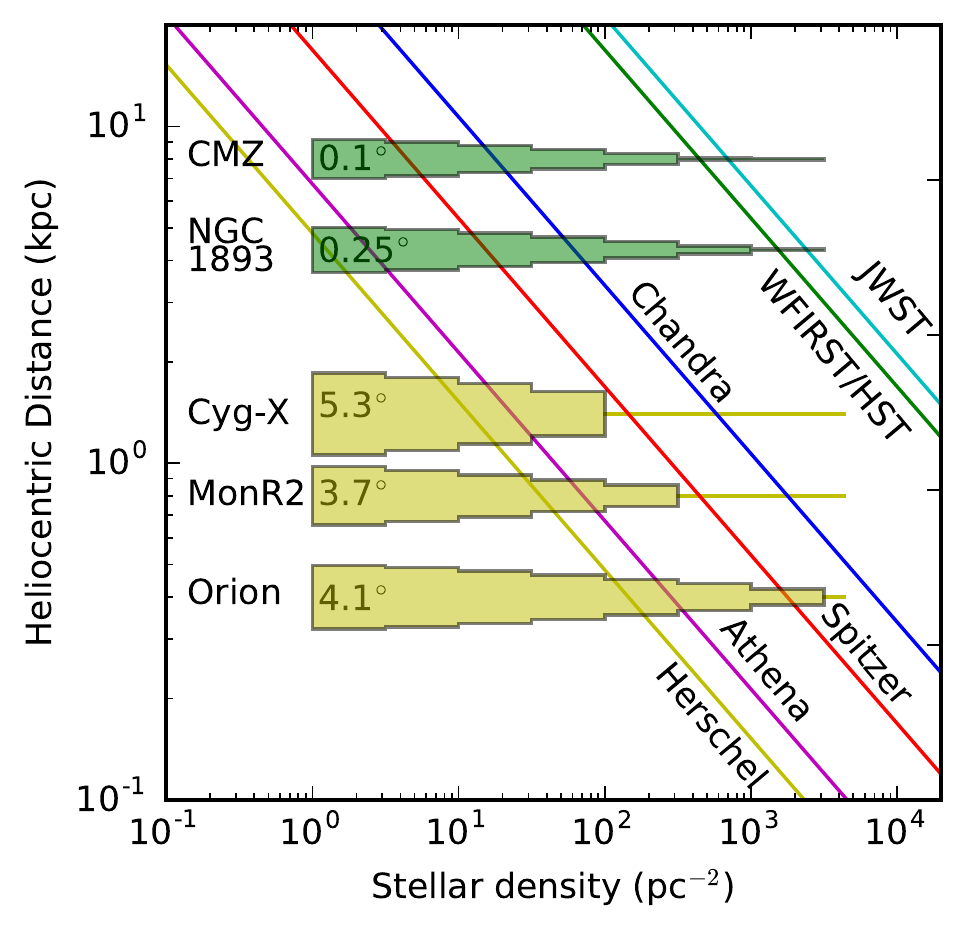}}
 \end{minipage}\hfill
  \begin{minipage}[c]{0.42 \textwidth}
  \vskip -.3 in
\caption{Distance vs range of stellar densities for select Milky Way regions. The widths of the wedges scale with the log of the solid angle subtended for a given range of stellar density, with the total solid angle above 1~pc$^{-2}$ displayed \citep{2011ApJ...739...84G,2014AJ....148...11K,2016AJ....151....5M}. The CMZ and outer galaxy region NGC~1893 are from \citet{2018ApJ...853..171G,2008A&A...488..211C}, their  wedges are that of Orion scaled to their $\ge 1$~pc$^{-2}$ solid angles. The diagonal lines show the point source confusion limit for {\it JWST} and {\it Spitzer} at 4.5~$\mu$m, {\it WFIRST/HST} at 1.6-2~$\mu$m, {\it Herschel} at 100~$\mu$m,  {\it Chandra}/ACIS, and {\it Athena} WFI.
}
\end{minipage}
\label{fig:environments}
\end{figure}


\noindent
{\bf Mid-IR imaging:} The detection of dusty disks and envelopes at 3-24~$\mu$m can be used to identify young stars and characterize their evolutionary states \citep{2009ApJS..184...18G,2009ApJS..181..321E, 2012AJ....144..192M}.  X-ray observations can mitigate incompleteness in the centers of clusters, where the bright IR nebulosity  hides stars  \citep{2016AJ....151....5M}. {\it JWST} will supply high resolution observations of distant clouds ($> 5$~kpc) and dense sub-regions in nearby clouds ($< 5$ ~kpc).  Wide field mapping with $\sim 1''$ resolution in the mid-IR is needed for surveys of clouds at 1-5 kpc.   

\noindent
{\bf Far-IR \& sub-mm imaging:} Imaging at 50-1000~$\mu$m can identify the youngest protostars \citep{2013ApJ...767...36S} as well as measure the SEDs and luminosities of protostars \citep{2016ApJS..224....5F}. Observatories with higher sensitivities and angular resolutions than {\it Herschel} will extend observations to denser and more distant regions (Fig.~1). With less precision, luminosities of protostars can also be extrapolated from mid-IR data obtained with {\it JWST} or {\it SOFIA} \citep{2012AJ....144...31K}.

\noindent
{\bf Near-IR imaging:} Observations at 1-2.5~$\mu$m are essential for measuring the properties of the photospheres of pre-ms stars and using proper motions to both identify members and map dynamical motions. Wide field imaging is needed for surveys of clouds in the nearest 1.5 kpc, while more distant regions require AO on 8-30 meter telescopes \citep[e.g.,][] {2019A&A...622A.149G,2019ApJ...870...44H}. Wide field, multi-epoch, sub-arcsecond 2~$\mu$m surveys - such as {\it WFIRST} - will allow proper motion studies across molecular clouds (also see white paper by Stauffer et al.).


\noindent
{\bf Near-IR spectroscopy:} 1-2.5~$\mu$m spectra are needed to measure the masses and ages of pre-ms stars, determine accretion rates, and determine radial velocities \citep[e.g.,][]{2016ApJ...818...59D,2017ApJ...845..105D,2016ARA&A..54..135H}. Multi-object spectrometers on 4-8~m telescopes combined with IFUs on 8-30~m telescopes for the centers of dense clusters are needed to efficiently obtain spectra.

\noindent
{\bf X-ray imaging:} Young stars can be identified through X-ray detections of their highly active coronae \citep{2005ApJS..160..401P}. Such observations can efficiently survey embedded pre-ms stars with and without disks, although with a lack of sensitivity to the lowest mass stars \citep{2015ApJ...802...60K}. While continued imaging with {\it Chandra} and {\it XMM} will extend the spatial extent and depth of existing studies, anticipated improvements in sensitivity with missions such as {\it Athena} will enable both mapping of entire clouds in the nearest 1.5 kpc and studies of more distant regions (Fig.~1). 

\noindent
{\bf Interferometery at cm wavelengths:} The VLBA can measure motions of stars with non-thermal emission from active coronae \citep{2017ApJ...834..142K}, next generation centimeter-wave facility capable of 1-10 mas resolution will enable proper motion measurements of thermal emitters. 

\noindent
{\bf Environmental diversity:} Observations of spectral lines, dust continuum, and polarized dust emission will differentiate environments through their gravitational, thermal, turbulent, and magnetic energy densities and structures (see Kauffmann et al. and Gutermuth et al. whitepapers).

\noindent
{\bf Support for ground based data analysis and simulations:} Although space-based facilities provide funding to analyze data, enhanced funding for ground-based observations and theory are necessary to support near-IR observations and produce simulations to interpret the data.


\vskip 0.1 in 

\centerline{\bf \large Application 1: The Star Formation Rate in Diverse Environments}

\vskip 0.1 in

\begin{figure}[t!]
\begin{minipage}[c]{0.5\textwidth}\center{\includegraphics[width = 4.4 in, trim = 0 50 50 20 0]{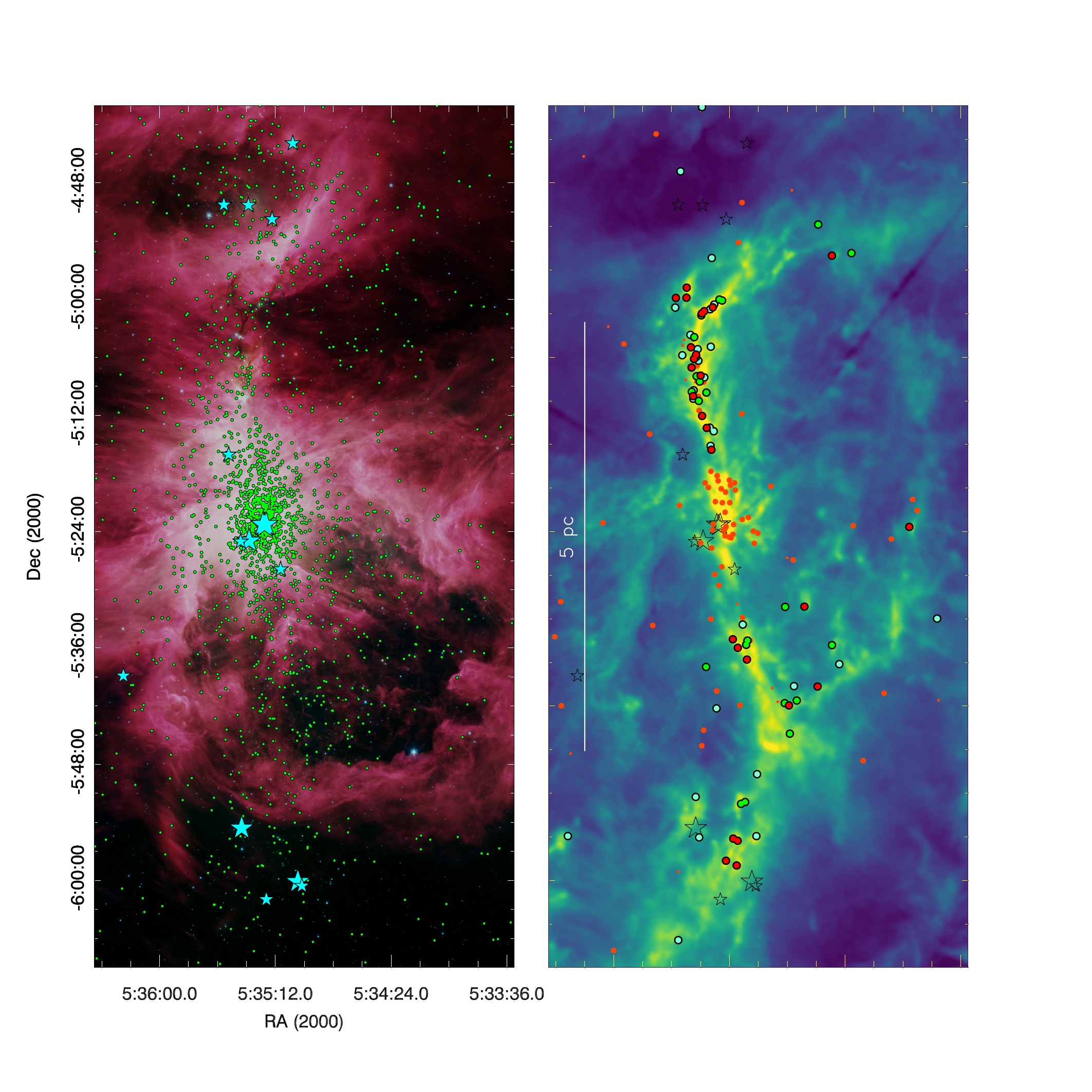}}
 \end{minipage}\hfill
  \begin{minipage}[c]{0.305 \textwidth}
\caption{Results of {\it Spitzer} and {\it Herschel} surveys of the integral shaped filament hosting the Orion Nebula Cluster (ONC) - part of the 50~pc long Orion~A cloud. The left panel shows the {\it Spitzer}/IRAC image with the locations of pre-ms stars with disks overlaid \citep{2012AJ....144..192M,2016AJ....151....5M}. The right panel shows the {\it Herschel} derived column density map with the locations of protostars marked as circles \citep{2016ApJS..224....5F,2016A&A...590A...2S}. The star shaped symbols are the optically visible O-B3 stars from \citet{1994A&A...289..101B}.} 
\end{minipage}
\label{fig:isf}
\end{figure}

The star formation rate (SFR) measures the conversion of interstellar gas into stars and plays a fundamental role in baryon cycles within galaxies. Surveys with {\it Spitzer} and {\it Herschel} have provided the means to determine the SFR in molecular clouds by directly mapping the distribution and properties of low-mass stars and protostars (Fig.~2). These data show that the SFR per surface area varies by two orders of magnitude across molecular clouds within 1 kpc of the Sun, primarily as a function of gas column density \citep[e.g.,][]{2011ApJ...739...84G}. These  variations motivate systematic measurements of the SFR as a function of the natal environmental conditions, i.e. the density and temperature of molecular gas, intensity of the radiation field, and strength of the magnetic field. Empirical correlations between SFR and environment will lead to a deeper understanding of the  mechanisms that regulate star formation and are critical for extrapolating our knowledge of local star formation to more extreme environments in our galaxy and others.

Although the integrated emission from star-forming regions, typically driven by the most massive stars, is used to characterize galactic scales, synthetic observations of hydrodynamic simulations show that such indicators  break down for individual regions \citep{2017ApJ...849....2K}. This underscores the importance of pushing the completeness of young star and protostar number counts beyond the local Gould Belt clouds to more distant regions. For example, source counts from ALMA observations of the central molecular zone (CMZ) currently show orders of magnitude lower SFRs for a given gas column density compared to nearby regions \citep{2018ApJ...853..171G}.   

Measurements of  SFRs have advanced due to the availability of IR surveys of entire clouds; these measurements have been combined with gas column density maps to determine the star forming laws for nearby clouds \citep{2010ApJ...723.1019H,2011ApJ...739...84G,2013ApJ...778..133L}. Future work will build on the approaches established for nearby clouds: 
i.) taking the density of all detected pre-ms and protostars (from X-ray and IR surveys) and dividing by the estimated duration of star formation (from IR spectroscopy), this has the disadvantage that YSOs  migrate (Gutermuth et al. 2011),  ii.) taking the density of protostars and dividing by a typical protostellar lifetime (Lada et al. 2013), which has the disadvantage that protostellar lifetimes  may systematically vary, and  iii.) using the luminosities of the protostars measured from combined {\it Spitzer}, {\it Herschel} and sub-mm data to measure their instantaneous accretion rates \citep[e.g.,][]{2017ApJ...840...69F}.   

{\it JWST} will enable measurements of the SFR using low-mass young stars in more distant, extreme environments; these observations will be complemented by more sensitive X-ray and far-IR observations as well as near-IR data from 8-30 meter telescopes (Fig.~1). Observations of the SFR go hand in hand with theoretical and numerical models, which are essential to calibrate the emission as a function of stellar environment, resolution and completeness as well to build a predictive framework for the SFR as a function of local and global galactic physical conditions.
	


\vskip 0.1 in
\centerline{\bf \large Application 2: Searching for Variations in the IMF with Environment}
\vskip 0.1 in

The detection of systematic variations in the IMF would have implications for both the physical basis of the mass function and the utilization of high mass stars as tracers of star formation in galaxies \citep{2010ARA&A..48..339B}. Establishing the presence and nature of such variations is best done in  molecular clouds and young clusters where dynamical processes and stellar evolution have not significantly altered the stellar mass distribution and the physical conditions of the birth environment have not been completely erased (see also Calzetti et al. \& Gutermuth et al. whitepapers).  


Evidence of spatial variations in the IMF has been driven by surveys for low-mass (proto)-stars \citep[e.g., ][]{2012ApJ...745..131K,2014ApJ...782L...1E}. \citet{2012ApJ...752...59H,2013ApJ...764..114H} find variations in the ratio of high- to low-mass stars between the Orion Nebula Cluster (ONC), located at the northern end of the Orion A cloud, and the network of small clusters and groups  at the southern end of the cloud. These results suggest  that the higher mass stars form preferentially in regions of higher gas and stellar density, and that the IMF of the southern, lower density part of the cloud is top light compared to the ONC. \citet{2019ApJ...870...44H} use proper motions from {\it HST}/WFC3 data to identify members of the young Arches cluster near the galactic center. Combining WFC3 photometry with near-IR spectroscopy of the most massive stars, they find that the cluster IMF down to  1.8~M$_{\odot}$ is unusually shallow. They propose that the IMF of the CMZ and galactic center is top heavy.  

\begin{figure}[t!]
\begin{minipage}[c]{0.5\textwidth}\center{\includegraphics[width = 3.4 in, trim =  0 30 20 20]{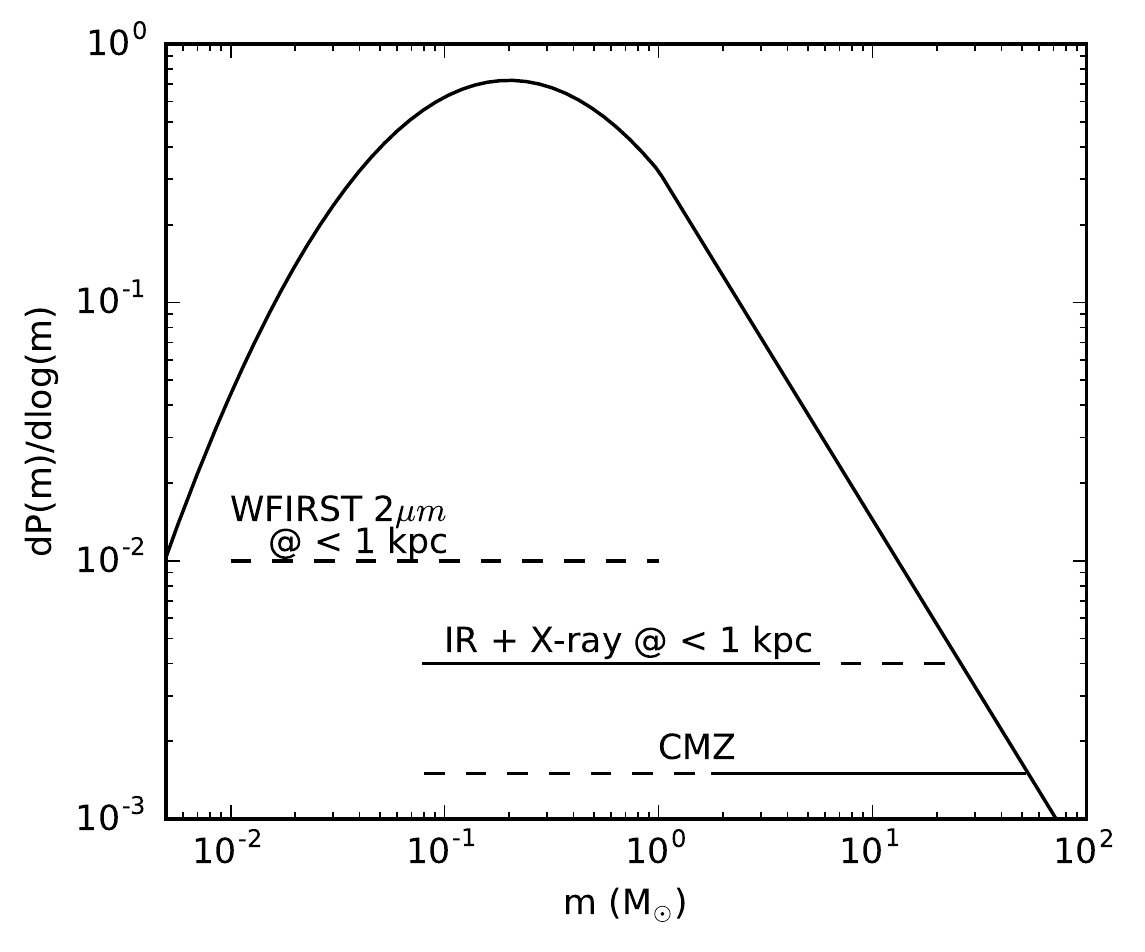}}
 \end{minipage}\hfill
  \begin{minipage}[c]{0.4\textwidth}
  \vskip -0.2 in
\caption{Comparison of the mass ranges of proposed studies compared to the IMF of \citet{2003PASP..115..763C}. The solid lines show current capabilities while the dashed lines give desired limits.  Limitations on the high-mass ends are driven by the small sample sizes while the low-mass end limitations are driven by sensitivity and confusion limits.  In the case of the {\it WFIRST}, proper motions will be key for identifying faint members, particularly in lower stellar density regions. }
\end{minipage}
\label{fig:imf}
\end{figure}



To further establish the presence of systematic IMF variations in our galaxy, future investigations must proceed in two directions: surveys of the IMF in lower stellar density populations within 1.5 kpc and the study of dense, extreme clusters at larger distances (Fig.~3). The former requires the means to reliably identify members and then determine their spectral types over wide fields. Embedded young stars can be identified in X-ray and IR surveys and characterized by multi-object spectrographs on 2-4 meter telescopes for regions within 500 pc and 4-8 meter telescopes for the richer 0.5-1.5 kpc regions. These searches can also detect  variations in the IMF below the hydrogen burning limit. Distant clusters require both the detection and characterization of low-mass stars in regions with significant extinction. Here, {\it JWST}, {\it WFIRST}, and future generation X-ray telescopes with $\le 1''$ resolution will provide the opportunity to identify pre-ms stars, while spectrographs on {\it JWST} and  8-30 meter telescopes with AO will be able to determine spectral types. In both directions, low spectral resolution spectrographs ($\sim 300$) will be particularly efficient for obtaining spectral types of stars around the characteristic mass of the IMF \citep[e.g.,][]{2008ApJ...685..313P}.


Tightening systematic and statistical observational uncertainties in turn increases constraints on theoretical models. As a result of large uncertainties,  the origin of the IMF and how it varies with physical conditions remains debated, although there is consensus that stellar feedback is the key ingredient responsible for limiting IMF variation with environment \citep{2014prpl.conf...53O}. Observations of the IMF for a variety of environments is essential to test the relative influence of gas temperature, stellar feedback and turbulence on the IMF.  At the lowest stellar masses, debate continues about whether brown dwarfs and very low-mass stars are formed like their more massive counterparts, i.e., in dense cores \citep{2004ApJ...617..559P}, or from the gravitational fragmentation of massive accretion disks \citep{2016ARA&A..54..271K}.  From a modeling perspective, the problem of star  formation in clusters involves a large dynamic range in density ($>10$ orders of magnitude) and size scale ($> 5$ orders of magnitude) in addition to multi-physics treatments, which requires super-computing facilities. Continued computational infrastructure support is necessary to extend the stellar mass limits and achieve statistical significance ($N_*> 10^3$) in star cluster simulations, while fully exploring the relationship between environmental conditions and the IMF.



\vskip 0.1 in
\centerline{\bf \large Application 3: The Formation  of Bound Clusters}
\label{sec:motion}
\vskip 0.1 in

The motions of young, low-mass stars trace the processes that both assemble young clusters and drive their evolution through gas dispersal. Observations of the 3D motions can test whether embedded clusters form  through cold collapse, hierarchical merging or  oscillating filaments \citep{2009ApJ...697.1103T,2009ApJ...697.1020P,2016A&A...590A...2S}. They can also trace evolution through the  gas dispersal to understand the environmental conditions necessary to form bound clusters from young gas-dominated clusters \citep{2010MNRAS.404..721M,2019ApJ...871...46K}.  

Radial velocity observations can be used to compare the motions of stars to radial velocities measured in clouds \citep[Fig.~4, e.g.,][]{2015ApJ...799..136F,2016A&A...589A..80H}. Deployment of multi-object, near-IR spectrometers with  high spectral resolution ($> 10,000$) on 8-30 meter telescopes will provide the means to efficiently measure velocities in   clusters beyond 1~kpc. Multiplicity can dominate the radial velocities, hence multi-epoch observations are key \citep[e.g.,][]{2019ApJ...871...46K}. Proper motions of embedded stars are needed to measure 3D velocities and track the motions of stars along filamentary clouds \citep[Fig.~4, e.g.,][]{2018AJ....156...84K,2019ApJ...870...32K}. Multi-epoch observations with {\it HST}, {\it JWST}, and ground-based instruments can measure these motions. {\it WFIRST}, particularly if equipped with a 2~$\mu$m filter that will both increase the sensitivity to deeply embedded objects and  lower contributions from scattered light that can limit positional accuracy, will provide an unprecedented facility for measuring proper motions over cloud scales.



A great deal of work has been devoted to exploring both the earliest stages of forming star clusters or the dynamics of  clusters once most of the gas has dispersed, but do not simulate gas dispersal \citep[e.g., ][]{2007MNRAS.380.1589B,2009ApJ...704L.124O,2012MNRAS.419.3115B,2012MNRAS.425..450M,2018MNRAS.473.2372K,2019MNRAS.483.4999F}. However, due to computational constraints, far less work has focused on the evolution of young stars from birth through the dispersal of their natal cloud. To date, efforts have either failed to resolve small scale fragmentation and multiplicity or to realistically simulate the gas dispersal \citep[e.g.,][]{2014MNRAS.439.1765K,2015ApJ...809..187S,2015MNRAS.451..987D,2018MNRAS.476.5341F}. Future work modelling this transitional phase, including both feedback effects and the formation of individual stars/stellar systems, is essential to connect the kinematics of clusters with their birth conditions and understand the conditions needed to form bound clusters. 
\vskip 0.2 in
\begin{figure}[b!]
\begin{minipage}[c]{0.5\textwidth}\center{\includegraphics[width = 4.5 in, trim =  0 0 0 20]{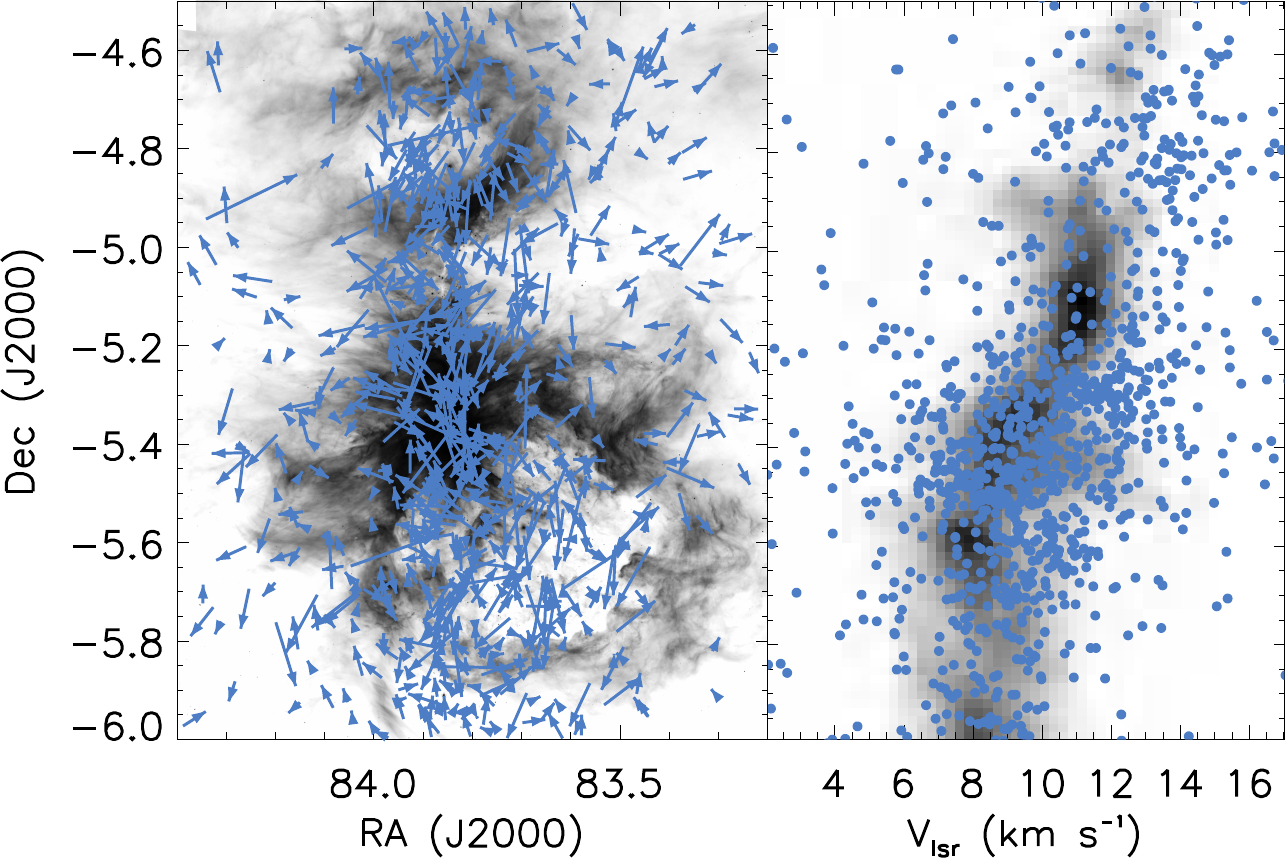}}
 \end{minipage}\hfill
  \begin{minipage}[c]{0.28\textwidth}
\vskip -.48 in\caption{Kinematics of the ONC. Left: \textit{Gaia} proper motions in the cluster reference frame against the 8 $\mu$m Spitzer image of \citet{2012AJ....144..192M}. {\it WFIRST} will measure proper motions of embedded stars invisible to {\it Gaia}. Right: radial velocities from near-IR APOGEE survey \citep{2018AJ....156...84K}, with the greyscale background showing the gas kinematics traced by $^{13}$CO \citep{1987ApJ...312L..45B}.}
\end{minipage}
\label{fig:motion}
\end{figure}



\bibliography{refs}

\end{document}